\documentclass[%
reprint,
superscriptaddress,
showpacs,preprintnumbers,
amsmath,amssymb,
aps,
]{revtex4-1}
\usepackage{graphicx}% Include figure files
\usepackage{dcolumn}% Align table columns on decimal point
\usepackage{xfrac}
\usepackage{bm}% bold math
\usepackage{color}% text color
\usepackage[colorinlistoftodos]{todonotes} %package for comments
\usepackage[export]{adjustbox}
\usepackage{hyperref}% add hypertext capabilities
\usepackage{tabularx}
\newcommand{\NIO}{Na$_2$IrO$_3$}
\newcommand{\LIO}{$\alpha$-Li$_2$IrO$_3$}

\newcommand{\ALIO}{Ag$_{3}$LiIr$_2$O$_6$}
\newcommand{\ALSO}{Ag$_{3}$LiSn$_2$O$_6$}
\newcommand{\HLIO}{H$_{3}$LiIr$_2$O$_6$}

\newcommand{\C}{$^\circ$C}
\begin{document}
%\preprint{APS/123-QED}

\title{Thermodynamic Evidence of Proximity to a Kitaev Spin-Liquid in \ALIO}
%Optimizing the Intra-layer Exchange Interactions by Altering the Inter-layer Bonds in the Kitaev Magnet \ALIO}
%\title{Role of the inter-layer Bonds in a New Kitaev Magnet \ALIO}
% Force line breaks with \\

\author{Faranak~Bahrami}
\affiliation{Department of Physics, Boston College, Chestnut Hill, MA 02467, USA}
\author{William~Lafargue-Dit-Hauret}
\affiliation{Univ Rennes, CNRS, ISCR (Institut des Sciences Chimiques de Rennes) – UMR 6226, F-35000 Rennes, France}
\affiliation{Physique Th\'{e}orique des Mat\'{e}riaux, CESAM, Universit\'{e} de Li\`{e}ge, B-4000 Sart Tilman, Belgium}
\author{Oleg~I.~Lebedev}
\affiliation{Laboratoire CRISMAT, ENSICAEN-CNRS UMR6508, 14050 Caen, France}
\author{Roman~Movshovich}
\affiliation{MPA-CMMS, Los Alamos National Laboratory, Los Alamos, NM 87545}
\author{Hung-Yu~Yang}
\affiliation{Department of Physics, Boston College, Chestnut Hill, MA 02467, USA}
\author{David~Broido}
\affiliation{Department of Physics, Boston College, Chestnut Hill, MA 02467, USA}
\author{Xavier~Rocquefelte}
\affiliation{Univ Rennes, CNRS, ISCR (Institut des Sciences Chimiques de Rennes) – UMR 6226, F-35000 Rennes, France}
\author{Fazel~Tafti}
\affiliation{Department of Physics, Boston College, Chestnut Hill, MA 02467, USA}
\email{fazel.tafti@bc.edu}

\date{\today}% It is always \today, today,

\begin{abstract}
Kitaev magnets are materials with bond-dependent Ising interactions between localized spins on a honeycomb lattice.
Such interactions could lead to a quantum spin-liquid (QSL) ground state at zero temperature.
Recent theoretical studies suggest two potential signatures of a QSL at finite temperatures, namely a scaling behavior of thermodynamic quantities in the presence of quenched disorder, and a two-step release of the magnetic entropy.
Here, we present both signatures in \ALIO\ which is synthesized from \LIO\ by replacing the inter-layer Li atoms with Ag atoms.
In addition, the DC susceptibility data confirm absence of a long-range order, and the AC susceptibility data rule out a spin-glass transition.
These observations suggest a closer proximity to the QSL in \ALIO\ compared to its parent compound \LIO\ that orders at 15~K.
We discuss an enhanced spin-orbit coupling due to a mixing between silver $d$ and oxygen $p$ orbitals as a potential underlying mechanism.
\end{abstract}

%\pacs{71.20.Eh, ?????}% PACS, the Physics and Astronomy
\maketitle

%\tableofcontents

\begin{figure}
\includegraphics[width=0.45\textwidth]{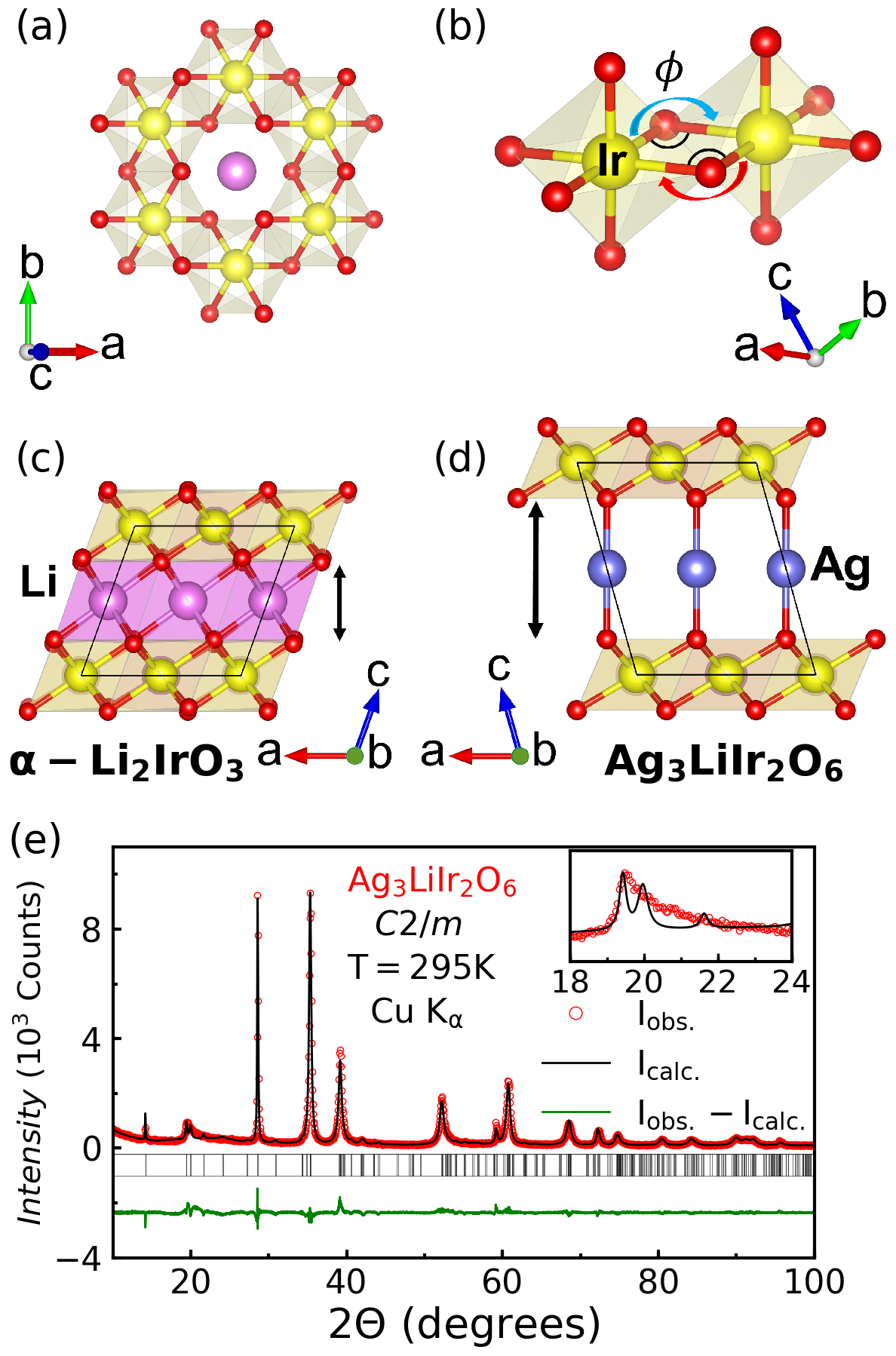}
\caption{\label{BOND}
(a) Honeycomb lattice of edge-sharing IrO$_6$ octahedra in both \LIO\ and \ALIO.
(b) Heisenberg exchange paths between neighboring octahedra.
(c) Octahedral coordination of Li atoms between the layers of \LIO.
(d) Linear (dumbbell) coordination of Ag atoms between the layers of \ALIO\ which leads to increasing the inter-layer separation.
(e) Rietveld analysis with a magnified view of the Warren line shape due to stacking faults (See also Figs.~S1 and S3~\cite{suppmatt}).
}
\end{figure}

%%%%%%%%%%%%%%%%%%%%%%%%%%%%%%%%%%%%%%%%%%%%%%%%%%%%%%%%%%%%%%%%%%%%%
%% Intoduction
%%%%%%%%%%%%%%%%%%%%%%%%%%%%%%%%%%%%%%%%%%%%%%%%%%%%%%%%%%%%%%%%%%%%%
%\section{Introduction}
%%
An exciting frontier in condensed matter physics is to design materials where the spin degrees of freedom avoid a magnetically ordered ground state despite strong exchange interactions.
Such compounds release the spin entropy by forming a quantum entangled ground state known as the quantum spin-liquid (QSL)~\cite{knolle_field_2019,takagi_concept_2019,savary_quantum_2016,winter_challenges_2016}.
Among various proposals for a QSL, the Kitaev model is especially appealing because it is exactly solvable and can be engineered in real materials~\cite{kitaev_anyons_2006,jackeli_mott_2009}.
The model consists of bond-dependent Ising interactions between localized $S=1/2$ spins on a honeycomb lattice, $\mathcal{H}_K=-\sum K_{\gamma}\mathbf{S_i}^{\gamma}\mathbf{S_j}^{\gamma}$~\cite{kitaev_anyons_2006,knolle_field_2019}.
The ground state is analytically solved by fractionalizing the spin-1/2 operators ($\mathbf{S_i}$) into itinerant and localized Majorana fermions~\cite{kitaev_anyons_2006,hermanns_physics_2018}.
Recent Monte Carlo (MC) simulations suggest that by decreasing temperature, the two types of Majoranas undergo two successive cross-overs~\cite{nasu_thermal_2015,yamaji_clues_2016}.
First, at a higher temperature $T_H$, the itinerant Majoranas form coherent bands.
Second, at a lower temperature $T_L$, the localized Majoranas form Z$_2$ gauge fluxes aligned on all hexagons.
Evidence of such behavior is reported in layered iridium oxides, \LIO\ and \NIO, with a honeycomb network of edge-sharing IrO$_6$ octahedra (Fig.~\ref{BOND}a) where Ir$^{4+}$ assumes a $J_{\mathrm{eff}}=1/2$ state due to strong spin-orbit coupling (SOC)~\cite{mehlawat_heat_2017}.
However, both compounds exhibit long-range antiferromagnetic (AFM) ordering and fail to reach a QSL ground state~\cite{choi_spin_2012,williams_incommensurate_2016,choi_spin_2019}.
Thus, a complete model Hamiltonian for the honeycomb iridates must include non-Kitaev interactions:
\begin{equation}
\label{Kit}
\mathcal{H}=\sum\limits_{\langle i,j \rangle \in {\alpha\beta(\gamma)}} \left[ -K_{\gamma}S_i^{\gamma}S_j^{\gamma} + J \textbf{S}_i  \cdot \textbf{S}_j+ \Gamma \left( S_i^{\alpha}S_j^{\beta} + S_i^{\beta}S_j^{\alpha}\right) \right]
\end{equation}
where the Kitaev term ($K$) favors QSL, the Heisenberg term ($J$) favors AFM order, and the off-diagonal exchange term ($\Gamma$) controls details of the magnetic order~\cite{jackeli_mott_2009,rau_generic_2014}.
Both \LIO\ and \NIO\ seem to be closer to the Heisenberg limit ($J>K$) despite evidence of a strong Kitaev interaction~\cite{banerjee_proximate_2016,hwan_chun_direct_2015}.
%%

%%%%%%%%%%%%%%%%%%%%%%%%%%%% FIGURE1 %%%%%%%%%%%%%%%%%%%%%%%%%%%%%%%%
%%%%%%%%%%%%%%%%%%%%%%%%%%%%%%%%%%%%%%%%%%%%%%%%%%%%%%%%%%%%%%%%%%%%%
%%
Recently, two approaches have been taken to bring the candidate materials closer to the Kitaev limit.
The first approach was to bring the Ir-O-Ir bond angles closer to 90$^\circ$ and maximize a destructive quantum interference between the Heisenberg interactions across each pair of super-exchange paths~\cite{jackeli_mott_2009} (Fig.~\ref{BOND}b).
This idea led to the discovery of $\alpha$-RuCl$_3$~\cite{plumb_2014} where the AFM order occurs at $T_N=7$~K~\cite{cao_low-temperature_2016} smaller than $T_N=15$~K in iridates.
The second approach was to induce a random bond disorder within the honeycomb layers which is achieved in \HLIO\ due to hydrogen intercalation and a heavy stacking disorder~\cite{bette_solution_2017,kitagawa_spinorbital-entangled_2018,knolle_bond-disordered_2019}.
Here, we present a third approach based on modifying the inter-layer bonds.
We replace the Li atoms between the layers of \LIO\ (Fig.\ref{BOND}c) with Ag atoms to produce \ALIO\ (Fig.\ref{BOND}d).
The honeycomb layers of \ALIO\ are identical to those of its parent compound but the chemical bonds between the layers are modified.
The inter-layer Li atoms in \LIO\ are octahedrally coordinated with six oxygens, three on top and three at the bottom, whereas the Ag atoms in \ALIO\ are linearly coordinated with two oxygens (Fig.~\ref{BOND}c,d).
The weaker O-Ag-O dumbbell bonds result in a 30\% increase of the inter-layer separation.
Our experiments reveal three thermodynamic signatures that suggest \ALIO\ is closer to the Kitaev limit than its parent compound \LIO.
First, the AFM peak in the magnetic susceptibility of \LIO\ at 15~K is absent in \ALIO.
Second, a scaling behavior is observed in the AC susceptibility over three decades of $T/H$ consistent with a random singlet scenario in QSL candidates~\cite{kimchi_scaling_2018}.
Third, a two-step release of the magnetic entropy at $T_H=75$~K and $T_L=13$~K is observed consistent with recent MC simulations~\cite{nasu_thermal_2015,yamaji_clues_2016}.
%%

%%%%%%%%%%%%%%%%%%%%%%%%%%%%%%%%%%%%%%%%%%%%%%%%%%%%%%%%%%%%%%%%%%%%%
%% Experimental
%%%%%%%%%%%%%%%%%%%%%%%%%%%%%%%%%%%%%%%%%%%%%%%%%%%%%%%%%%%%%%%%%%%%%
%\section{\label{exp}Methods}
%%
Polycrystalline samples of \ALIO\ were prepared via a topotactic reaction at 350~\C\ for 24~h according to
\begin{equation}
\mathrm{2Li_2IrO_3 + 3AgNO_3 \rightarrow Ag_3LiIr_2O_6 + 3LiNO_3}
\end{equation}
The precursor \LIO\ was synthesized following prior reports~\cite{mehlawat_heat_2017}.
We also synthesized the non-magnetic \ALSO\ using a similar procedure, and used it as a phonon analogue of \ALIO\ in the heat capacity analysis.
Powder X-ray diffraction (PXRD) was performed using a Bruker D8 ECO instrument.
The FullProf suite~\cite{rodriguez-carvajal_recent_1993} was used for the Rietveld refinement.
Magnetization and heat capacity were measured using Quantum Design MPMS3 and PPMS Dynacool, respectively.
%%

%%%%%%%%%%%%%%%%%%%%%%%%%%%%%%%%%%%%%%%%%%%%%%%%%%%%%%%%%%%%%%%%%%%%%
%% Results and Discussion
%%%%%%%%%%%%%%%%%%%%%%%%%%%%%%%%%%%%%%%%%%%%%%%%%%%%%%%%%%%%%%%%%%%%%

%\section{Results and discussion}

%%%%%%%%%%%%%%%%%%%%%%%%%%%%%%%%%%%%%%%%%%%%%%%%%%%%%%%%%%%%%%%%%%%%%
%% XRD
%%%%%%%%%%%%%%%%%%%%%%%%%%%%%%%%%%%%%%%%%%%%%%%%%%%%%%%%%%%%%%%%%%%%%
%\subsection{\label{crystallography}X-ray Crystallography}
%%
\emph{Structure--} Figure~\ref{BOND}e shows the PXRD pattern of \ALIO\ with a Rietveld refinement in the same space group ($C2/m$) as its parent compound \LIO~\cite{suppmatt,omalley_structure_2008}.
Although a prior work has suggested the space group $R\bar{3}m$~\cite{todorova_agrho2_2011}, a recent structural analysis of the material agrees with our solution~\cite{bette_crystal_2019}.
The asymmetric broadening (Warren line shape) of the peaks between 18 and 24$^\circ$ in the inset of Fig.~\ref{BOND}e is commonly observed in the layered honeycomb structures~\cite{roudebush_structure_2013,abramchuk_crystal_2018,abramchuk_cu2iro3:_2017}.
It is analyzed in the Supplemental Fig.~S1 and gives at least 5\% of stacking disorder~\cite{suppmatt}.
Our Rietveld refinement~\cite{suppmatt} shows relatively small Debye-Waller factors for the Ag atoms~\cite{suppmatt} corresponding to well-defined Ag-O bonds unlike the H-O bonds in \HLIO\ where the region of stacking faults must be excluded to obtain a reasonable refinement~\cite{bette_solution_2017}.
Thus, the in-plane bond randomness in \HLIO~\cite{knolle_bond-disordered_2019} is negligible in \ALIO.
To gain further confidence on the reported oxygen positions and Ir-O-Ir bond angles, we subjected the crystallographic unit cell to a geometric optimization in the VASP code~\cite{kresse_efficiency_1996,suppmatt}.
The results in Table~\ref{T1} (and Fig.~S2) show an excellent agreement between the experimental and theoretical bond distances and angles.
We performed the same analysis on \LIO\ and found comparable Ir-O-Ir bond angles between the two compounds (Table~\ref{T1}).
Thus, the cancellation between opposite Heisenberg exchange paths in Fig.~\ref{BOND}b must be comparable between \ALIO\ and \LIO.
However, their magnetic behavior is different as discussed next.
%%

%%%%%%%%%%%%%%%%%%%%%%%%%%%%% TABLE1 %%%%%%%%%%%%%%%%%%%%%%%%%%%%%%%%
\begin{table}
  \caption{\label{T1}Experimental and theoretical values of bond lengths and angles in \ALIO\ and \LIO.
  }
  \begin{tabular}{c|cc}
  \hline
  \hline
  \multicolumn{3}{c}{ \ALIO }                                     \\
  \hline
                     &    Experimental        &  Theoretical      \\
   Ir1-O1-Ir1        &    $96.5(3)^\circ$    &  $97.54(0)^\circ$ \\
   Ir1-O2-Ir1        &    $96.9(6)^\circ$    &  $97.66(0)^\circ$ \\
   Ir1-O1            &    $2.043(9)$~\AA      &  $1.988(0)^\circ$       \\
   Ir1-O2            &    $2.046(5)$~\AA      &  $1.990(0)^\circ$       \\
   \hline
   \hline
   \multicolumn{3}{c}{ \LIO }                                     \\
   \hline
                     &    Experimental       &  Theoretical       \\
   Ir1-O1-Ir1        &    $94.7(5)^\circ$      &  $94.42(0)^\circ$  \\
   Ir1-O2-Ir1        &    $95.3(8)^\circ$      &  $94.56(0)^\circ$  \\
   Ir1-O1            &    $2.015(13)$~\AA      &  $2.003(0)^\circ$       \\
   Ir1-O2            &    $2.080(19)$~\AA      &  $2.010(0)^\circ$       \\
   \hline
   \hline
   \end{tabular}
\end{table}
%%%%%%%%%%%%%%%%%%%%%%%%%%%%%%%%%%%%%%%%%%%%%%%%%%%%%%%%%%%%%%%%%%%%%

%%%%%%%%%%%%%%%%%%%%%%%%%%%%%%%%%%%%%%%%%%%%%%%%%%%%%%%%%%%%%%%%%%%%%
%% Magnetization
%%%%%%%%%%%%%%%%%%%%%%%%%%%%%%%%%%%%%%%%%%%%%%%%%%%%%%%%%%%%%%%%%%%%%
%\subsection{\label{magnetization}Magnetic Susceptibility}
%%%%%%%%%%%%%%%%%%%%%%%%%%%% FIGURE3 %%%%%%%%%%%%%%%%%%%%%%%%%%%%%%%%
\begin{figure*}
\includegraphics[width=\textwidth]{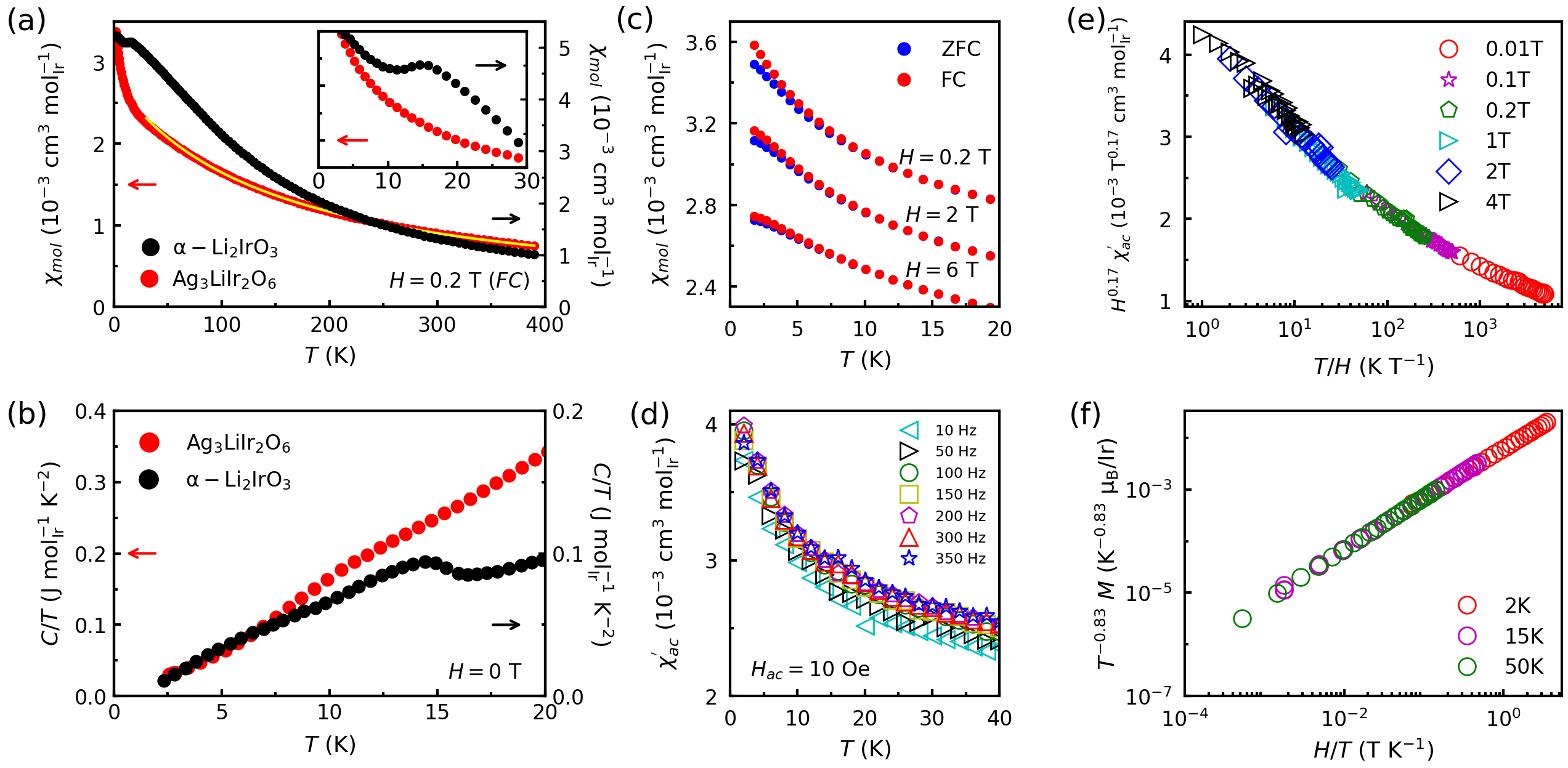}
\caption{\label{CW}
(a) DC Magnetic susceptibility as a function of temperature in \ALIO\ (red) and \LIO\ (black) with a magnified view below 30~K in the inset.
The yellow line is a Curie-Weiss fit.
(b) Heat capacity per mole Ir as a function of temperature in \ALIO\ (red) and \LIO\ (black data from ref.~\cite{singh_relevance_2012}).
(c) A small splitting in the DC susceptibility data under ZFC and FC conditions appears below 10~K. It disappears at higher fields. The curves are slightly shifted for visbility.
(d) The real part of the AC susceptibility $\chi'_{ac}$ as a function of temperature. 
(e) Data collapse for $H^{\alpha}\chi'_{ac}$ as a function of $T/H$ on a semi-log scale with $\alpha=0.17$.
(f) Data collapse for $T^{1-\alpha}M$ as a function of $H/T$ on a log-log scale.
}
\end{figure*}
%%%%%%%%%%%%%%%%%%%%%%%%%%%%%%%%%%%%%%%%%%%%%%%%%%%%%%%%%%%%%%%%%%%%%
%%
\emph{Magnetism--}
Figure~\ref{CW}a shows that the peak at $T_N=15$~K in the magnetic susceptibility of \LIO\ due the AFM ordering is absent in \ALIO.
Similarly, Fig.~\ref{CW}b confirms the absence of a peak in the heat capacity of \ALIO\, unlike the peak at 15~K in \LIO.
However, a slight change of slope is discernible in \ALIO\ at $T_L=13$~K.
These observations suggest that the second-order AFM transition in \LIO\ is replaced by a cross-over in \ALIO.
The yellow line in Fig.~\ref{CW}a is a fit to the expression $\chi=\chi_0+\frac{C}{T-\Theta_{\textrm{cw}}}$ which yields a Curie-Weiss temperature $\Theta_{\textrm{cw}}=-142$~K and a magnetic moment $\mu=1.79\,\mu_B$ comparable to the reported values in \LIO\ ($-105$~K, $1.83\,\mu_B$)~\cite{mehlawat_heat_2017,singh_relevance_2012}.
This is consistent with the similar bond angles in Table~\ref{T1} and confirms a comparable strength of the Heisenberg exchange interaction in both compounds.
A small splitting between the zero-field-cooled (ZFC) and field-cooled (FC) curves is observed below 10~K (Fig.~\ref{CW}c) that suggests a trace of spin glass-like freezing.
As seen in Fig.~\ref{CW}c and d, this splitting is only 3\% of the total magnetization, vanishes at higher fields, and does not produce a peak in the AC susceptibility.
Thus, it originates from a minority of frozen spins (quenched disorder) while the majority of the system remains in a paramagnetic QSL state.
A universal behavior among QSL materials with quenched disorder is a data collapse as reported in H$_3$LiIr$_2$O$_6$, LiZn$_2$Mo$_3$O$_8$, ZnCu$_3$OH$_6$C$_{12}$, and Cu$_2$IrO$_3$~\cite{kimchi_scaling_2018,kitagawa_spinorbital-entangled_2018,kenney_coexistence_2019,choi_exotic_2019}.
The data collapse results from a subset of random singlets induced by a small amount of disorder within either a spin-liquid or a valence-bond-solid (VBS) ground state~\cite{kimchi_scaling_2018}.
Figure~\ref{CW}e presents a data collapse of $H^{0.17}\chi_{ac}$ as a function of $T/H$ over three decades of the scaling parameter.
Similarly, Fig.~\ref{CW}f shows a scaling of $T^{-0.83}M$ as a function of $H/T$.
These scaling analyses confirm the presence of random singlets in \ALIO\ but cannot distinguish between a spin-liquid or a VBS ground state.
%%

%\subsection{\label{heat}Heat Capacity}
%%%%%%%%%%%%%%%%%%%%%%%%%%%% FIGURE 4 %%%%%%%%%%%%%%%%%%%%%%%%%%%%%%%%
\begin{figure}
\includegraphics[width=0.44\textwidth]{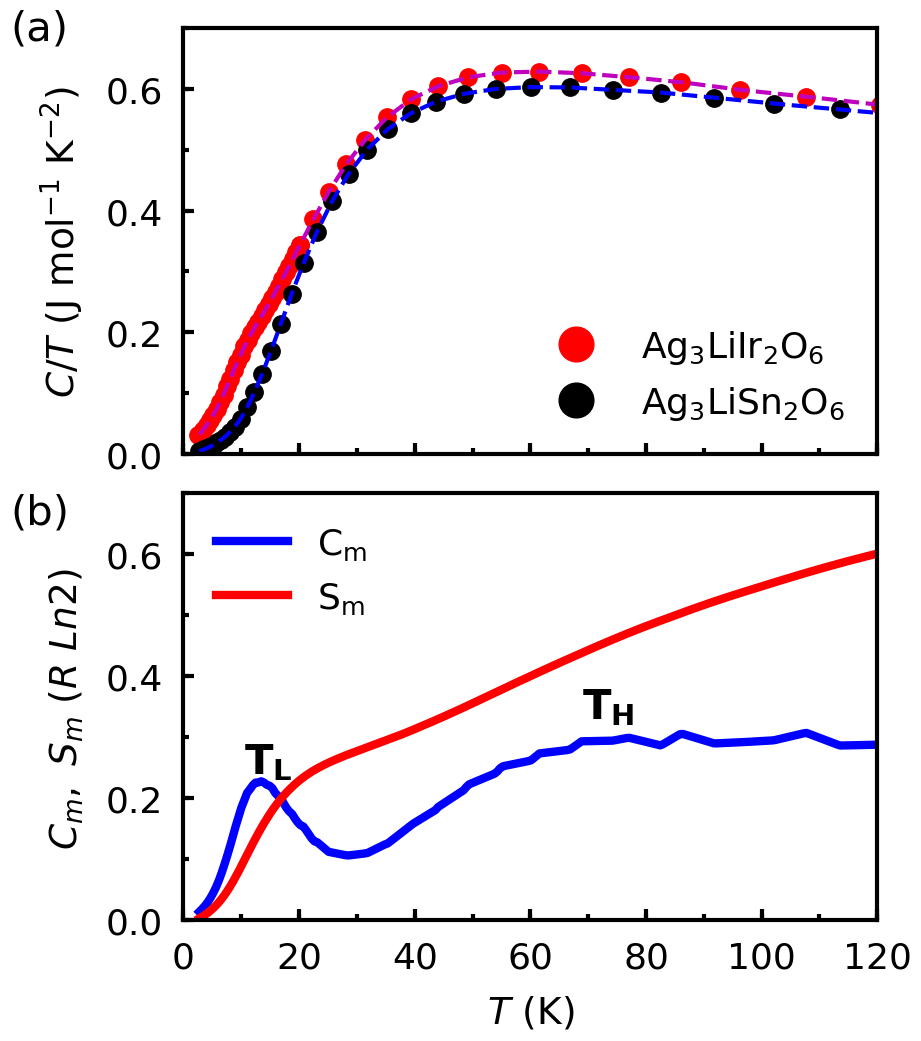}
\caption{\label{SC}
(a) Heat capacity ($C/T$ per mole Ir or Sn) plotted as a function of temperature in \ALIO\ and its lattice model \ALSO.
(b) Magnetic heat capacity ($C_m$) and entropy ($S_m$) plotted in units of $R\ln(2)$ as a function of temperature.
Two broad features are revealed at $T_H\approx75$~K and $T_L=13$~K.
}
\end{figure}
%%%%%%%%%%%%%%%%%%%%%%%%%%%%%%%%%%%%%%%%%%%%%%%%%%%%%%%%%%%%%%%%%%%%%
\emph{Heat capacity--} As mentioned in the introduction, the MC simulations suggest that a Kitaev magnet releases the spin entropy in two successive cross-overs at a higher ($T_H$) and a lower ($T_L$) temperature~\cite{nasu_thermal_2015}.
In 3D, for example in a hyperhoneycomb lattice, these cross-overs turn into phase transitions~\cite{kimchi_three-dimensional_2014,nasu_finite-temperature_2014}.
Figure~\ref{SC}a presents $C/T$ (per mole Ir or Sn) as a function of temperature in \ALIO\ and \ALSO, where the stannate is used to subtract the phonon background from the iridate.
The resulting magnetic heat capacity $C_m$ is plotted as a function of $T$ in Fig.~\ref{SC}b and used to calculate the magnetic entropy via $S_m=\int{\frac{C_m}{T}dT}$ that reveals a two step structure.
The first step is broad and corresponds to the broad hump at $T_H\approx75$~K in $C_m$.
The second step is better resolved and corresponds to the peak at $T_L=13$~K in $C_m$.
Neither of these features are sharp, i.e. they are more likely to be cross-overs instead of second-order AFM transitions.
This behavior is qualitatively consistent with the MC simulations~\cite{nasu_thermal_2015,yamaji_clues_2016}; however, two deviations from the theory must be pointed out.
(a) according to theory, the entropy release at each step must be $\frac{1}{2}R\ln(2)$, but we observe $60\%$ of this value.
A similar observation is reported in the parent compound, \LIO, and the quantitative disagreement is attributed to the phonon background subtraction~\cite{mehlawat_heat_2017}.
It is possible that \ALSO\ is not a perfect lattice model.
(b) ideally, the ratio of $T_L/T_H$ should be less than $0.03$ for a Kitaev spin-liquid~\cite{yamaji_clues_2016,nasu_thermal_2015}, but $T_L/T_H = 0.17$ in \ALIO, similar to both \LIO\ and \NIO~\cite{mehlawat_heat_2017}.
Note that the MC simulations were performed on an ideal system with purely Kitaev interactions.
Because the real candidate materials have additional non-Kitaev interactions (Eq.~\ref{Kit}), it is expected to find mild deviations from the ideal theoretical results.
%%

%%%%%%%%%%%%%%%%%%%%%%%%%%%%%%%%%%%%%%%%%%%%%%%%%%%%%%%%%%%%%%%%%%%%%
%% Discussion
%%%%%%%%%%%%%%%%%%%%%%%%%%%%%%%%%%%%%%%%%%%%%%%%%%%%%%%%%%%%%%%%%%%%%
%%%%%%%%%%%%%%%%%%%%%%%%%%%%% TABLE2 %%%%%%%%%%%%%%%%%%%%%%%%%%%%%%%%
\begin{table}
  \caption{\label{T2}Comparing the experimental values of the average Ir-O-Ir bond angle ($\phi$), Curie-Weiss temperature ($\Theta_{\textrm{cw}}$), inter-layer separation ($d$), and Ir--Ir distance between \ALIO\ and \LIO.%, and \NIO.
  The $c$-axis parameter and the monoclinic angle $\beta$ for \LIO\ are from the reference~\cite{omalley_structure_2008} and $\Theta_{\textrm{cw}}$ is from reference~\cite{mehlawat_heat_2017}.
  }
   \begin{tabular}{cccc}
   \hline
   \hline
                         &   \ALIO                &    \LIO          \\%&   \NIO            &  \HLIO \\
   \hline
   $\bar{\phi}$          &    $96.7^\circ$        &  $95.0^\circ$    \\%&   $98.7^\circ$    &  $99.5^\circ$ \\
   $\Theta_{cw}$         &    $-142$~K            &  $-105$~K        \\%&   $-123$~K        &   $105$~K\\
   $d=c\sin(\beta)$      &    $6.24$~\AA          &  $4.82$~\AA      \\%&   $5.31$~\AA      &  $4.54$~\AA\\
   Ir--Ir                &    $3.04$~\AA          &  $2.98$~\AA      \\%&   $3.13$~\AA    \\
   \hline
   \hline
   \end{tabular}
\end{table}
%%%%%%%%%%%%%%%%%%%%%%%%%%%%%%%%%%%%%%%%%%%%%%%%%%%%%%%%%%%%%%%%%%%%
%%%%%%%%%%%%%%%%%%%%%%%%%%%% FIGURE 4 %%%%%%%%%%%%%%%%%%%%%%%%%%%%%%%%
\begin{figure}
\includegraphics[width=0.44\textwidth]{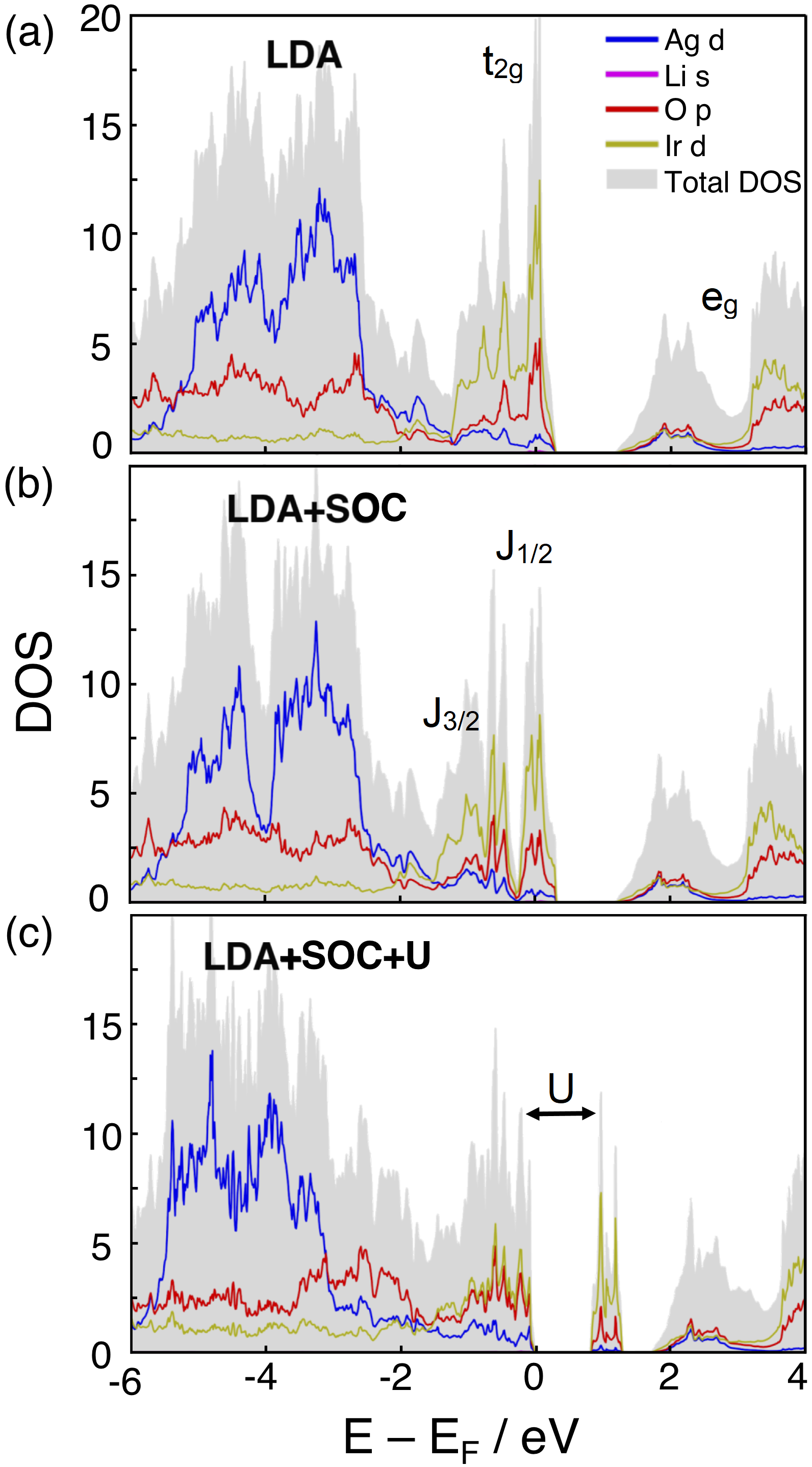}
\caption{\label{DOS}
Density of states calculated at three levels of DFT with (a) local density approximation (LDA), (b) LDA+SOC, and (c) LDA+SOC+$U$ where $U$ is the exchange potential.
}
\end{figure}
%%%%%%%%%%%%%%%%%%%%%%%%%%%%%%%%%%%%%%%%%%%%%%%%%%%%%%%%%%%%%%%%%%%%%
%%
\emph{Discussion--} At this point, it is instructive to compare the structural and magnetic parameters between \ALIO\ and \LIO\ (Table~\ref{T2}).
Due to a comparable bond angle $\phi$, the cancellation of Heisenberg interactions across the opposite Ir-O-Ir bonds in Fig.~\ref{BOND}b must be comparable in both compounds.
A comparison of $\Theta_{\textrm{cw}}$ and Ir--Ir distance suggests that the exchange coupling strength is also comparable in both compounds.
The main structural difference between the two materials is a 30\% larger inter-layer separation in \ALIO.
At first glance, an increased inter-layer separation may suggest increased magnetic fluctuations, hence a weaker AFM order.
However, the exchange interactions in iridate materials are highly anisotropic~\cite{sizyuk_importance_2014} and such an argument does not justify the complete suppression of the AFM order in \ALIO.
A more plausible explanation for the lack of AFM order comes from the density of states (DOS) calculations presented in Fig~\ref{DOS} where a finite weight of silver $4d$ orbitals is observed at the Fermi level $E_F$.
We present three levels of the DFT calculations following the prior work on \LIO~\cite{li_analysis_2015}.
First, a plain local density approximation (LDA) is presented in Fig.~\ref{DOS}a to show the $t_{2\textrm{g}}$ states just below $E_F$ and $e_\textrm{g}$ states above $E_F$.
Notice that the majority of Ag electrons (blue line) are between 2 and 4~eV below $E_F$; however, a small but finite contribution from silver $d$ orbitals is observed near $E_F$.
Second, by adding the spin-orbit coupling (LDA+SOC) in Fig.~\ref{DOS}b, the $t_{2\textrm{g}}$ levels are split into lower $J_\textrm{eff}=3/2$ and an upper $J_\textrm{eff}=1/2$ states.
Third, by adding an exchange potential (LDA+SOC+U) in Fig.~\ref{DOS}c, a gap is opened within the $J_\textrm{eff}=1/2$ states to separate the upper and lower Hubbard bands.
These results are identical to \LIO\ and consistent with the localized effective spin-1/2 Kitaev model~\cite{li_analysis_2015}.
The new finding is the finite weight of silver $4d$ orbitals at $E_F$ which remains unchanged between the LDA and LDA+SOC+U calculations, and suggests a $d$-$p$ orbital mixing between the Ag and O atoms.
Whereas the lithium $2s$ electrons in \LIO\ are transferred to oxygen $2p$ orbitals in an ionic bond, the silver $4d$ electrons in \ALIO\ are more extended and bonded to the oxygen $2p$ orbitals with a more covalent character.
As a result of such $d$-$p$ mixing, the SOC is effectively increased on the Ir-O-Ir exchange path within the honeycomb layers of \ALIO\ which enhances the Kitaev coupling.
We emphasize that despite comparable Ir-O-Ir bond angles between \LIO\ and \ALIO\ within the honeycomb layers (Table~\ref{T1}), the latter compound is closer to the Kitaev limit because of a stronger SOC mediated via the O-Ag-O bonds between the layers.
Thus, our work presents a new approach to optimizing the Kitaev magnets by tuning the inter-layer instead of intra-layer chemical bonds.
%%

% %%
% The absence of AFM ordering and the two-step release of entropy in \ALIO\ are consistent with a quantum paramagnetic ground state which could be either a spin-liquid or a VBS in the absence of disorder~\cite{savary_quantum_2016}.
% %
% A small amount of quenched disorder could randomize a subset of valence bonds within the quantum paramagnetic ground state~\cite{kimchi_scaling_2018}.
% %
% The scaling analyses presented in Fig.~\ref{CW}e,f are consistent with such a random singlet scenario within a quantum paramagnetic ground state.
% %%

%%%%%%%%%%%%%%%%%%%%%%%%%%%%%%%%%%%%%%%%%%%%%%%%%%%%%%%%%%%%%%%%%%%%%
%% Conclusions
%%%%%%%%%%%%%%%%%%%%%%%%%%%%%%%%%%%%%%%%%%%%%%%%%%%%%%%%%%%%%%%%%%%%%

%%%%%%%%%%%%%%%%%%%%%%%%%%%%%%%%%%%%%%%%%%%%%%%%%%%%%%%%%%%%%%%%%%%%%
%% Acknowledgements
%%%%%%%%%%%%%%%%%%%%%%%%%%%%%%%%%%%%%%%%%%%%%%%%%%%%%%%%%%%%%%%%%%%%%

\section*{ACKNOWLEDGMENTS}
We thank I.~Kimchi, Y.~Ran, N.~Perkins, and D.~Haskel for fruitful discussions.
The work at Boston College was supported by the National Science Foundation under award number DMR--1708929.
Work at Los Alamos was conducted under the auspices of the U.S. Department of Energy, Office of Basic Energy Sciences, Division of Materials Sciences and Engineering.
O.I.L acknowledges financial support from the "Agence Nationale de la Recherche" in the framework of the "Investissements d'avenir" program with the reference "ANR--11--EQPX--0020" for EELS data obtained using GIF Quantum.
This work was granted access to the HPC resources of [TGCC/CINES/IDRIS] under allocation 2017-A0010907682 made by GENCI.
%%

%%%%%%%%%%%%%%%%%%%%%%%%%%%%%%%%%%%%%%%%%%%%%%%%%%%%%%%%%%%%%%%%%%%%%
%% Appendix
%%%%%%%%%%%%%%%%%%%%%%%%%%%%%%%%%%%%%%%%%%%%%%%%%%%%%%%%%%%%%%%%%%%%%

% \appendix

% \section{\label{BK}\msr\ data and background subtraction}
% %%%%%%%%%%%%%%%%%%%%%%%%%%%%%%%%%%%%%%%%%%%%%%%%%%%%%%%% FIGURE 3 %%%%%%%%%%%%%%%%%%%%%%%%%%%%%%%%%%%%%%%%%%%%%%%%%%%%%%%%%
% \begin{figure}
% \includegraphics[width=0.45\textwidth]{MuSR_2}
% \caption{\label{MuSR_2}
% %
% Representative zero field (ZF) spectra obtained by the MuSR spectrometer at 15, 7, and 1.7~K from a sample inside helium exchange cryostat.
% %
% Solid lines are fits to Eq.~\ref{musrfit}.
% }
% \end{figure}
% %%%%%%%%%%%%%%%%%%%%%%%%%%%%%%%%%%%%%%%%%%%%%%%%%%%%%%%%%%%%%%%%%%%%%%%%%%%%%%%%%%%%%%%%%%%%%%%%%%%%%%%%%%%%%%%%%%%%%%%%%%%

%The \nocite command causes all entries in a bibliography to be printed out
%whether or not they are actually referenced in the text. This is appropriate
%for the sample file to show the different styles of references, but authors
%most likely will not want to use it.
%\nocite{*}

\bibliography{Bahrami_30oct2019}% Produces the bibliography via BibTeX.

\end{document}